# Design and Implementation of an Unmanned Vehicle using a GSM Network with Microcontrollers

Sourangsu Banerji,
Department of Electronics & Communication Engineering,
RCC-Institute of Information Technology, India

**ABSTRACT :** Now-a-days, a lot of research is being carried out in the development of USVs (Unmanned surface vehicles), UAVs (Unmanned Aerial Vehicles) etc. Now in case of USVs generally, we have seen that wireless controlled vehicles use RF circuits which suffer from many drawbacks such as limited working range, limited frequency range and limited control. Moreover shooting infrared outdoors on a bright sunny day is often problematic, since sunlight can interfere with the infrared signal. Use of a GSM network (in the form of a mobile phone, a cordless phone) for robotic control can overcome these limitations. It provides the advantages of robust control, working range as large as the coverage area of the service provider in comparison with that of an IR system, no interference with other controllers. This paper presents a Global System for Mobile Telecommunication (GSM) network based system which can be used to remotely send streams of 4 bit data for control of USVs. Furthermore, this paper describes the usage of the Dual Tone Multi-Frequency (DTMF) function of the phone, and builds a microcontroller based circuit to control the vehicle to demonstrate wireless data communication. Practical result obtained showed an appreciable degree of accuracy of the system and friendliness through the use of a microcontroller.

## I. INTRODUCTION

RF control (often abbreviated to R/C or simply RC) is the use of radio signals to remotely control a device. The term is used frequently to refer to the control of model vehicles from a handheld radio transmitter. Industrial, military, and scientific research organizations make use of radio-controlled vehicles as well. A remote control vehicle (RCV) is defined as any mobile device that is controlled by a means that does not restrict its motion with an origin external to the device. This is often a radio control device, cable between control and vehicle, or an infrared controller. A RCV is always controlled by a human and takes no positive action autonomously.

The IR system follows the line of site approach of actually pointing the remote at the device being controlled; this makes communication to be impossible over obstacles and barriers. Moreover since IR systems suffer from these problems so to overcome this; a signaling scheme utilizing voice frequency tones is employed. This scheme is known as Dual Tone Multi-Frequency (DTMF), Touch-Tone or simply tone dialing. As its acronym suggests, a valid DTMF signal is the sum of two tones, one from a low group (697-941Hz) and the other from a high group (1209-1633Hz) with each group containing four individual tones. DTMF signaling plays an important role in distributed communication systems such as multiuser mobile radio (Zarlink, 1983). It is natural in the two way radio environment since it slips nearly into the center of the voice spectrum, has excellent noise immunity and it has a highly integrated method of implementation currently available. It is directly compactable with telephone signals simplifying automatic phone batch system.

The development of silicon implemented switch capacitors, sample-1 filters, marks the current generation of DTMF receiver technology. Initially single chip band pass filters were combined with currently available decoders enabling a two chip receiver design. A further advance in integration has merged both functions onto a single chip allowing DTMF receivers to be realized in minimal space of a low cost. Most Nokia phones have F-Bus and M-Bus connections that can be used to connect a phone to Personal Computers (PC) or in this case, a microcontroller. The connection can be used for controlling just about all functions of the phone.

In this paper, phones using GSM network interfaced with a microcontroller is used to remotely control an unmanned robotic vehicle thus overcoming distance barrier problem and communication over obstacles with very minimal or no interference but is solely network dependent. We present the design and implementation of an unmanned vehicle (i.e. a robotic vehicle) consisting of a GSM network (a mobile phone), DTMF decoder, microcontroller and a motor driver. The transmitter is a handheld mobile phone. Ordinary low cost mobile phones like Nokia 1100 or even older versions of Nokia phones could be used effectively for this purpose. There is no special requirement on the part of the mobile phones both of which are used in the transmitter and receiver section respectively. A system overview is given in section II, the hardware design framework is discussed in section III, the circuit design; construction and working, design calculation are given in section IV, V and VI. The software design framework, experimental results and cost analysis are covered in section VII, VIII and IX. The applications and future scope is explained in the section X and XI. Lastly we conclude our paper.

## II. SYSTEM OVERVIEW

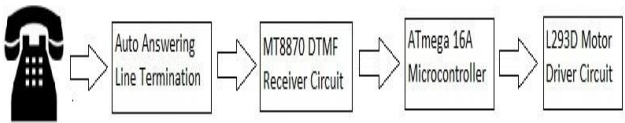

Figure1 DTMF Data Communication Architecture

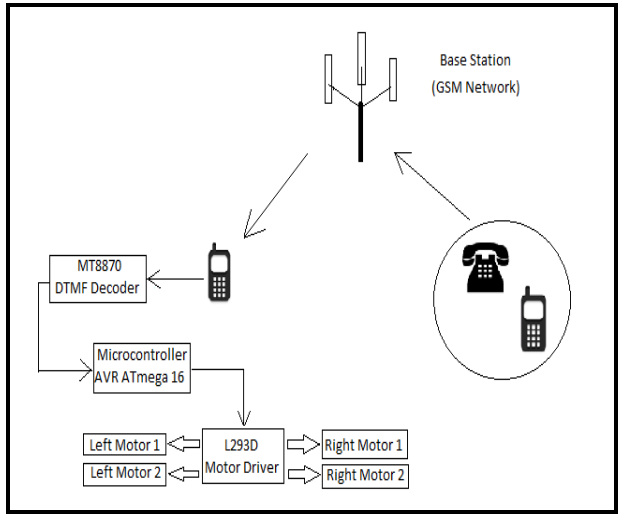

Figure 2 Functional block diagram of the system

The diagrams in fig.1 and fig.2 describe the overall system. Here the robotic vehicle is controlled by a mobile phone that makes a call to the GSM mobile phone attached to the robot. In the course of a call, if any button is pressed, a tone corresponding to the button pressed is heard at the other end of the call. This tone is called 'dual-tone multiple-frequency' (DTMF) tone. The robotic vehicle perceives this DTMF tone with the help of the phone stacked in the vehicle. The received tone is processed by the microcontroller with the help of DTMF decoder. The decoder decodes the DTMF tone into its equivalent binary digit and this binary number is sent to the microcontroller. The microcontroller is preprogrammed to take a decision for any given input and outputs its decision to motor drivers in order to drive the motors for forward or backward motion or turn left or right. The mobile that makes a call to the mobile phone stacked in the vehicle acts as a remote. So this simple robotic vehicle does not require the construction of receiver and transmitter units, reducing the overall circuit complexity.

## III. HARDWARE DESIGN FRAMEWORK

The blocks of the receiver model which is seen in fig.2 are explained in detail in this section:

### 3.1 DTMF DECODER

An MT8870 (Fig. 3) series DTMF decoder is used here. All types of the MT8870 series use digital counting techniques to detect and decode all the 16 DTMF tone pairs into a 4-bit code output. The built-in dial tone rejection circuit eliminates the need for pre-filtering. When the input signal given at pin 2 (IN-) in single-ended input configuration is recognized to be effective, the correct 4-bit decode signal of the DTMF tone is transferred to Q1 (pin 11) through Q4 (pin 14) outputs D0 through D3 outputs of the DTMF decoder (IC1) are connected to port pins of microcontroller.

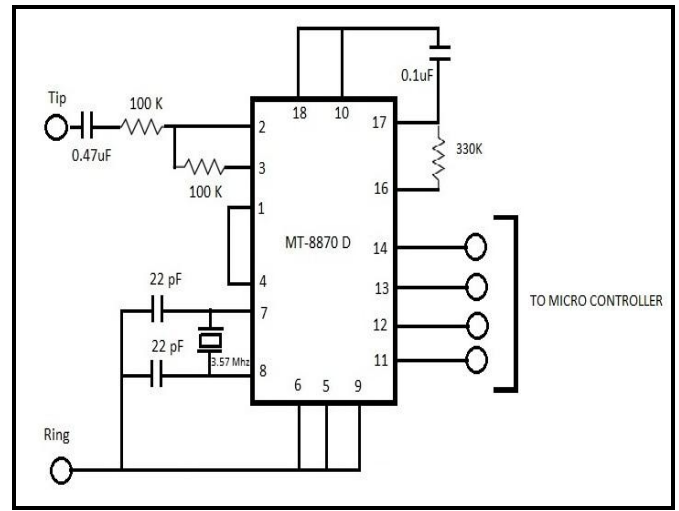

Figure 3

The MT8870 is a complete DTMF receiver, integrating both the band split filter and digital decoder functions. The filter section uses switch capacitor techniques for high and low group filters; the decoder uses digital counting techniques to detect and decode all 16 DTMF tone pairs into a 4 bit code. External component count is minimized by on chip provision of a differential input amplifier clock oscillator and latched three state bus interfaces. The functional description of the MT8870 is given in the following sections:

3.1.1 Filter Section:

Separation of the low group and high group tones is achieved by applying the DTMF signal to the input of the two sixth order switched capacitor band pass filter, the band width of which correspond to the low and high group frequencies. Each filter output is followed by a single order switched capacitor filter section which smoothens the signal prior to limiting; limiting is performed by high gain comparators which are provided with hysteresis to prevent detection of unwanted low level signals. The output of the comparators provides full rail logic swing at the frequency of the incoming DTMF signals.

3.1.2 Decoder Section:

Following the filter section is a decoder employing digital counting techniques to determine the frequencies of the incoming tones and to verify that they correspond to standard DTMF Frequencies. A complex averaging algorithm protects against tone simulation by extraneous signals such as voice while providing tolerance to small frequency deviations and variations. This averaging variation algorithm has been developed to ensure an optimum combination of immunity to talk off and tolerance to the presence of interfering frequencies (third tone) and noise. When the detector recognizes the presence of two valid tones (this is referred to as the signal condition, in some industry specifications) the early steering ($E_{st}$) output will go to an active state. Any subsequent loss of signal condition will course $E_{ST}$ to assume an inactive state.

3.1.3   Steering Circuit:

Before registration of a decoded tone pair, the receiver checks for a valid signal duration (referred to as character recognition condition) this check is performed by an external Resistance Capacitance (RC) time constant $E_{st}$. Logic high on $E_{st}$ causes collector voltage ($V_c$) to rise as the capacitor discharges. The time required to detect the presence of two valid tones top is function of the decode algorithm, the tone frequency and the previous state of the decode logic. $E_{ST}$ indicates and initiates an RC timing circuit. If both tones are present for the minimum guide time ($t_{CTP}$) which is determine by the external RC network, the DTMF signal is decoded and the resulting data is latched in the output register. The delay steering ($S_{tD}$) output is raised and indicates that new data is available. The time required to receive a valid DTMF signal ($T_{rec}$) is equal to the sum of time to detect the presence of valid DTMF signals ($t_{DP}$) and guard time, tone present.

### 3.2 MOTOR DRIVER

The L293D is a quad, high-current, half-H driver designed to provide bidirectional drive currents of up to 600 mA at voltages from 4.5V to 36V. It makes it easier to drive the DC motors.

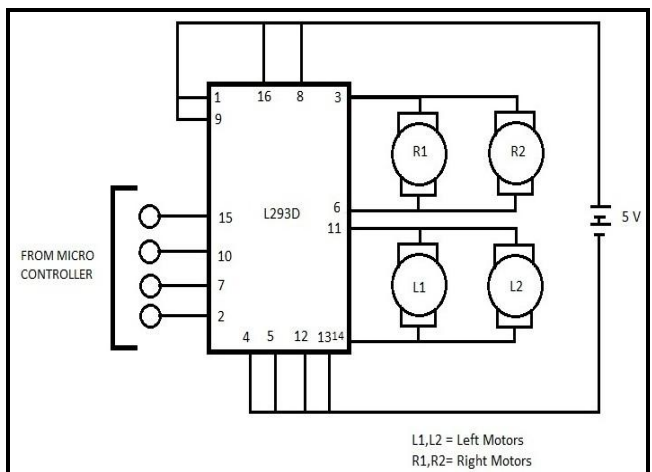

Figure 4

The L293D consists of four drivers. Pins IN1 through IN4 and OUT1 through OUT4 are input and output pins, respectively, of driver 1 through driver 4. Drivers 1 and 2, and drivers 3 and 4 are enabled by enable pin 1 (EN1) and pin 9 (EN2), respectively. When enable input EN1 (pin 1) is high, drivers 1 and 2 are enabled and the outputs corresponding to their inputs are active. Similarly, enable input EN2 (pin 9) enables drivers 3 and 4 [4], [5].

### 3.3 MICROCONTROLLER

The ATmega16 is a low-power CMOS 8-bit microcontroller based on the AVR enhanced RISC architecture. By executing powerful instructions in a single clock cycle, the ATmega16 achieves throughputs approaching 1 MIPS per MHz allowing the system designer to optimize power consumption versus processing speed.

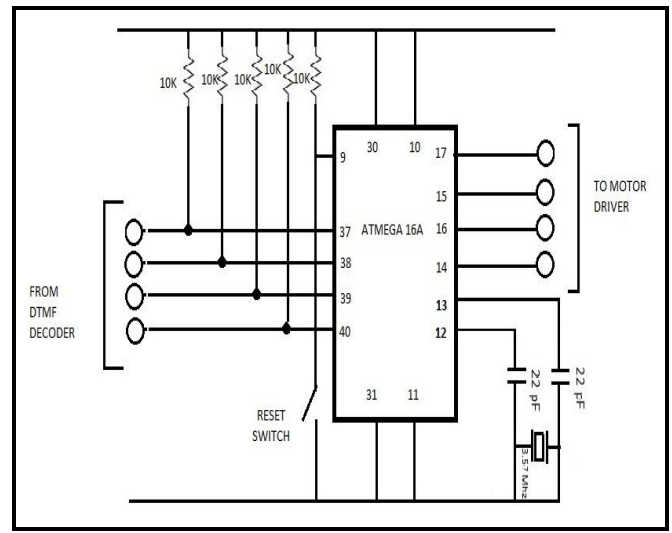

Figure 5

It provides the following features: 16 Kb of in-system programmable Flash program memory with read-while-write capabilities, 512 bytes of EEPROM, 1kB SRAM, 32 general-purpose input/output (I/O) lines and 32 general-purpose working registers. All the 32 registers are directly connected to the arithmetic logic unit, allowing two independent registers to be accessed in one single instruction executed in one clock cycle. The resulting architecture is more code-efficient. The on-chip Flash allows the program memory to be reprogrammed in-system or by a conventional nonvolatile memory programmer. By combining a versatile 8-bit CPU with Flash on a monolithic chip, the Atmel microcontroller is a powerful microcomputer, which provides a highly flexible and cost effective solution to many embedded control applications.

Outputs from port pins of the microcontroller are fed to inputs IN1 through IN4 respectively, to drive four geared DC motors. The microcontroller output is not sufficient to drive the DC motors, so current drivers are required for motor rotation.

### IV.   CIRCUIT DESIGN

In this section, the circuit diagram for the project is shown below

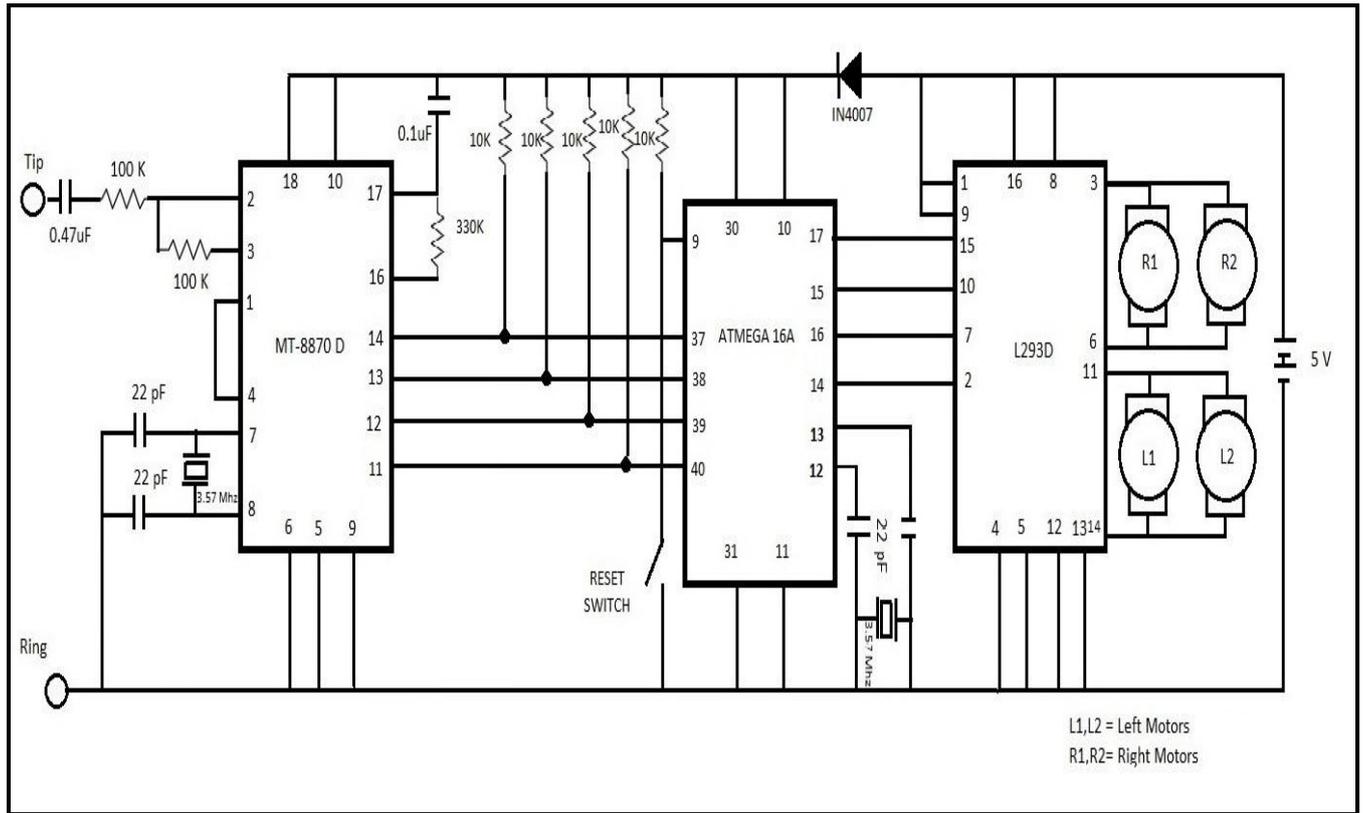

## V. CONSTRUCTION AND WORKING

While constructing any unmanned vehicle one major mechanical constraint is the number of motors being used. We can have either a two wheel drive or a four-wheel drive. In general we have seen that though four-wheel drive is more complex than two-wheel drive, but it also provides more torque and good control. The circuit configuration shown in the previous section was built on a simple breadboard.

Motors are fixed to the bottom of this steel chassis and the circuit is affixed firmly on top of this chassis. A cell phone is also mounted on the chassis. In the four-wheel drive system, the two motors on a side are controlled independently of each other. So a single L293D driver IC can drive the unmanned car. In order to control the vehicle, we need to make a call to the cell phone attached to the vehicle from any phone, which sends DTMF tones on pressing the numeric buttons. The cell phone in the robotic car is kept in 'auto answer' mode. So after a ring, the cell phone accepts the call. Now we may press any button on our mobile to perform actions. The DTMF tones thus produced are received by the cell phone in the robot. These tones are fed to the circuit by the headset of the cell phone. Referring to our circuit diagram given we must note that the ring and tip in the circuit refers to the two wires that we get when we split open the headphone wire of any GSM phone. If we cut open the ear phone we will find 3 basic wires coming out: 1. Red wire (headset right output-Ring), 2. Blue wire (headset left output-Tip) and 3.Copper wire (ground wire). These are generally known as TRS headphone wires. Special care should be taken in this regard as these wires are laminated and the lamination must be removed before the wires are connected to the MT8870 DTMF decoder. Otherwise the tones will not be received effectively by the decoder.

The MT8870 decodes the received tone and sends the equivalent binary number to the microcontroller. According to the program in the microcontroller, the robot starts moving. Port D ($PD_0$-$PD_7$) of Atmega16 has been designed as the output port of the microcontroller. We see that when key '2' is pressed on the mobile phone, the microcontroller outputs for forward motion. When we press key '8' on our mobile phone, the microcontroller outputs for reverse motion. When we press key '4' on our mobile phone, the microcontroller outputs for left direction motion. When we press key '6' on our mobile phone, the microcontroller outputs for right direction motion. Similarly when we press key '5' on our mobile phone, the microcontroller halts the vehicle. Five keys on the keypad are used for motion control of the unmanned car.

## VI. DESIGN CALCULATION

### 4.1 STEERING AND GUARD TIME CIRCUIT OPERATION FOR DECODER SECTION

$t_{GTP} < t_{GTA}$

$$t_{GTA} = (R_p C) \ln(\frac{V_{DD}}{V_{TST}}) \dots\dots\dots\dots\dots\dots\dots\dots (1)$$

$R_P = R_1 R_2 (R_1 + R_2) \dots\dots\dots\dots\dots\dots\dots\dots\dots\dots (2)$

Operating at $t_{GTP} > t_{GTA}$

$$t_{GTA} = (R_1 C) \ln(\frac{V_{DD}}{V_{TST}}) \dots\dots\dots\dots\dots\dots\dots\dots (3)$$

Typically $V_{st} = 0.5 V_{DD} \quad = \frac{5V}{2} = 2.5V \dots\dots\dots\dots (4)$

$t_{GTA} = (390 \times 10^3 \times 100 \times 10^{-9}) \ln 2 \dots\dots\dots\dots (5)$

given $\ln 2 = 0.693$,

$\simeq 390 \times 100 \times 10^{-6} \times 0.693 = 0.027027 \cong 0.03 sec \approx 3mSec$

$t_{GTA} = RC \ln\left(\frac{V_{DD}}{(V_{DD} - V_{TST})}\right)$ ................................. (6)

$t = (100 \times 10^{-9} \times 390 \times 10^3) \ln\left(\frac{5}{2.5}\right) = 0.03 sec \approx 3mSec.$

This is typical for DTMF. The actual value of $V_{DD}$ as measured with digital multi-meter is 4.9V compared to the 5V of the Microcontroller. The drop of 0.1V may be as a result of the drop across load resistances.

## 4.2 DIFFERENTIAL INPUT CONFIGURATION FOR MT8870

The input arrangement of the MT8870 provides a differential input operational amplifier as well as a bias source ($V_{ref}$) which is used to bias the input at mid rail. Provision is made for connection of a feedback resistor to the op-amp output (GS) for adjustment of gain.

$C_1 = C_2 = 10nF$
$R_1 = R_4 = R_5 = 100K\Omega$
$R_3 = 56K\Omega$

$R_2 = 37K\Omega$
$R_3 = 2\left(\frac{R_2 R_5}{R_2 + R_5}\right) = 2\left(\frac{39 \times 100}{39 + 100}\right) = 28.052K\Omega$

$2R_3 = 56K$

Voltage Gain $= \frac{R_5}{R_1} = \frac{100K\Omega}{100K\Omega} = 1$

Therefore the op-amp has unity voltage gain.

## 4.3 INPUT IMPEDENCE

$z_{indiff} = \sqrt[2]{R_1^2 + \left(\frac{1}{\omega c}\right)^2}$ ................................. (7)

$= \sqrt[2]{(100 \times 10^3)^2 + \left(\frac{1}{2\pi \times 685 \times 10 \times 10^{-9}}\right)^2} = 100K$

The design of DTMF receiving system can generally be broken down into three functional blocks. The first consideration is the interface to the transmission medium. This consists of few simple components to adequately configure the MT8870 input stage or may be as complex as some form of demodulation, multiplexing or analog switching system.

The second functional block is the DTMF receiver itself. This is the receiving system's interface parameter which can be optimized for specific signal condition delivered from the front end interface.

The third and most widely varying function is the output control logic. This control is responsible for handling system protocols and adaptively varying the tone receiver's parameter to adjust for changing signal conditions.

From the calculation above, the input voltage gain is unity. With the unity gain the MT8870 will accept maximum signal level of 1dBm into 600Ω. The lowest DTMF frequency that must be detected is approximately 685Hz.

Allowing 0.1dB of attenuation at 685Hz, the required input time constant is derived from:

$M(\omega)dB = 20 \log_{10} \frac{R_f}{R} + 20 \log_{10} \frac{\omega \tau}{((\omega \tau)^2 + 1)^{1/2}}$

Where $M(\omega) dB$ = Amplifier gain
$\omega$ in decibel = radian frequency
$\tau$ = input time constant

$-0.1 = 20 \log_{10} \frac{(2\pi)685\tau}{((2\pi*685\tau)^2 + 1)^{\frac{1}{2}}}$

$\tau = 1.52ms$

Choosing $R = 220K\Omega$ gives a high input impedance (440 less in the pass band)

$C = \frac{\tau}{R} = 6.9nF$

Using a standard value, C is chosen to be 10nF to achieve a unity gain in the pass band. $R_f$ is chosen to be equal to R that is; $R_f = R$, $R_a$ and $R_b$ are biasing resistors. The choice of $R_a$ is not critical and could be set at 68K say. Bias resistor $R_a$ adds a zero to the non-inverting path through the differential amplifier but has no effect on the inverting path.

## VII. SOFTWARE DESIGN FRAMEWORK

The software is written in 'C' language and compiled using AVR Studio 6.0 'C' compiler. The source program is converted into hex code by the compiler. We burned this hex code into ATmega16 AVR microcontroller. Now, we will first describe the algorithm of the code structure, follow it up a control flow diagram and implement the algorithm in actual C code.

### 7.1 ALGORITHM

In this section we are going to discuss the working algorithm which we have used in the construction of the unmanned vehicle:

*Step 1->* include the register name defined specifically for ATmega16 and also declare the variable.
*Step 2->* set port A as the input and port D as the output.
*Step 3->* the program will run forever by using 'while' loop.
*Step 4->* under 'while' loop, read port A and test the received input using 'switch' statement.

1. If 2 is pressed on the keypad, both the left and right motors move forward.
2. If 8 is pressed on the keypad, both the left and right motors moves backward.
3. If 4 is pressed on the keypad, the left motor halts and the right motor moves forward.
4. If 6 is pressed on the keypad, the right motor halts and the left motor moves forward.
5. If 5 is pressed on the keypad both the motors stop and the car comes to a stop.

*Step 5->* the corresponding data will output at port D after testing and conditioning of the received data.

This was how we maneuvered our unmanned vehicle using our handheld mobile phone.

### 7.2 CONTROL FLOW DIAGRAM

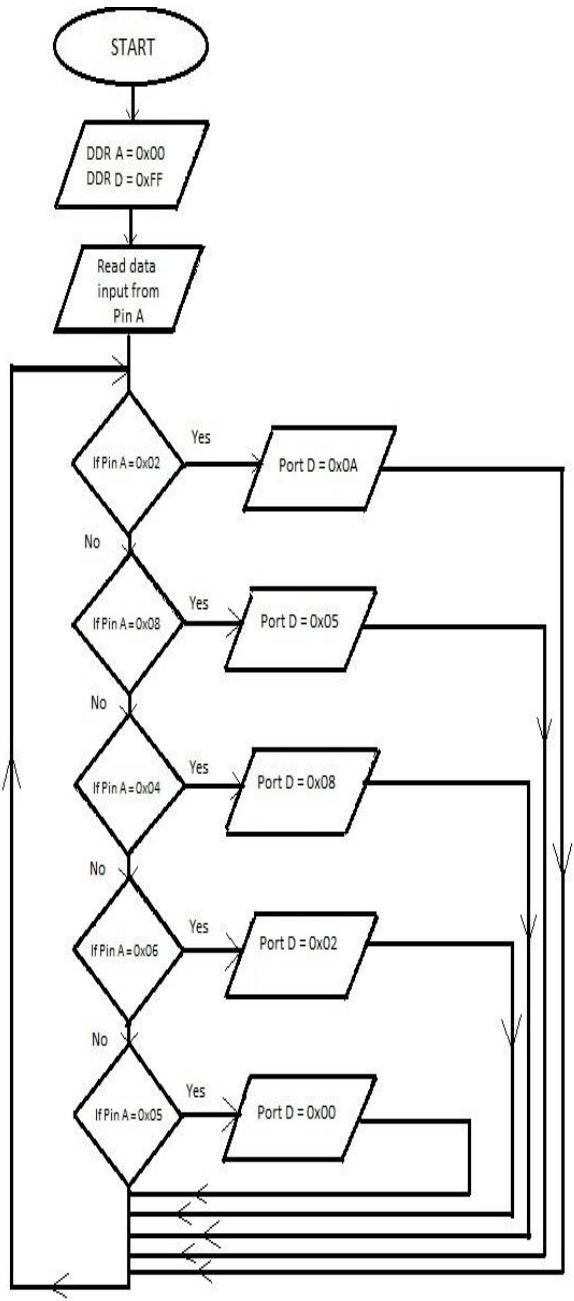

### 7.3 PROGRAM CODE

Here is the actual C code implementation:

```
#include <avr/io.h>   //standard I/O functions for ATMega16
int main(void)
{
  unsigned int k;
  DDRA= 0x00;    // All pins of PORTA assigned as INPUT
  DDRD=0xFF;     // All pins of PORTD assigned as OUTPUT
  PORTD=0x00;    // All pins made to exhibit LOW state initially
  while (1)      // Infinite Loop
  {
    k=PINA;      //Read Input from PINA
    switch (k)
    {
      case 0x02: //BOT Moves forward
      {
        PORTD=0x0A; // Both Motors in Forward Direction
        break;
      }
      case 0x08: // BOT Moves backward
      {
        PORTD=0x05; // Both Motors in Backward Direction
        break;
      }
      case 0x04: //BOT Moves Left
      {
        PORTD=0x08; // RM-Forward and LM-Stop
        break;
      }
      case 0x06: // BOT Moves Right
      {
        PORTD=0x02; // LM-Forward and RM-Stop
        break;
      }
      case 0x05: // BOT Stops
      {
        PORTD=0x00; //Both left and right motors halt
        break;
      }
    }
  }
}
```

## VIII. EXPERIMENTAL RESULTS

We measured experimental values of the frequency of sinusoidal waves for DTMF and voltage level at the output pins of MT8870 and Atmega16. These values are juxtaposed with the theoretical values in the Table 1 and 2 respectively. Table 3 shows the HEX reading obtained from output pins of MT8870 and Atmega 16.

TABLE 1
Frequency Readings

| KEY | LOWER FREQUENCY (Hz) | | HIGHER FREQUENCY(Hz) | |
|---|---|---|---|---|
| | TH. | EXP. | TH. | EXP. |
| 2 | 697 | 672 | 1336 | 1320 |
| 4 | 770 | 731 | 1209 | 1201 |
| 6 | 770 | 731 | 1477 | 1475 |
| 8 | 852 | 855 | 1336 | 1322 |
| 5 | 770 | 735 | 1336 | 1325 |

TABLE 2
Voltage Readings

| LOGIC LEVEL | OUPUT VOLTAGE OF MT8870 | | OUTPUT VOLTAGE OF MICROCONTROLLER | | OUTPUT VOLTAGE OF L293D | |
|---|---|---|---|---|---|---|
| | TH. | EXP. | TH. | EXP. | TH. | EXP. |
| LOW | 0.03 | 0.09 | 0 | 0.01 | 0 | 0.07 |
| HIGH | 4.97 | 4.80 | 5 | 4.2 | 5 | 4.93 |

TABLE 3
Hex readings and decisions taken

| KEY PRESSED | OUTPUT OF MT8870 | INPUT OF MICRO-CONTROLLER | OUTPUT OF MICRO-CONTROLLER | DECISION TAKEN |
|---|---|---|---|---|
| 2 | 0010 | 0010 | 0x0A | Forward |
| 4 | 0100 | 0100 | 0x08 | Left turn |
| 6 | 0110 | 0110 | 0x02 | Right turn |
| 8 | 1000 | 1000 | 0x05 | Backward |
| 5 | 0101 | 0101 | 0x00 | Stop |

TH. = Theoretical Value, EXP. = Experimental Value

## IX. COST ANALYSIS

A comparative cost analysis is an integral part of any project that is carried out. In this section, we try to provide an approximate cost estimate of the project. The components used here can be brought from any of the online electronics stores which can be found on the internet. For this project we purchased the components i.e. AVR Atmega 16, L293D IC etc from Mouser Electronics (India Centre) and the rest could be found in any college laboratory. We have listed the components and their prices in the following table:

TABLE 4
Cost of the Components

| Components | Quantity | Cost |
|---|---|---|
| Atmega 16 | 1 | $3.40 |
| MT8870 I.C. | 1 | $1.95 |
| L293D I.C. | 1 | $1.50 |
| IN4007 Rectifier diode | 1 | College Lab |
| 100kilo-ohm Resistor | 2 | College Lab |
| 330kilo-ohm Resistor | 1 | College Lab |
| 10kilo-ohm Resistor | 4 | College Lab |
| 0.47 uF | 1 | College Lab |
| 22pF | 4 | College Lab |
| 0.1uF | 1 | College Lab |
| Crystal (3.57MHz) | 1 | $0.15 |
| Crystal(12MHz) | 1 | $0.25 |
| Geared Motors | 4 | $12 |
| Breadboard | 1 | College Lab |
| Battery(6V) | 1 | $4.50 |
| | Total | $23.75 |

From the table above we can see that this unmanned vehicle can be made with components which are cheap and very readily available in the market. The circuit was designed keeping in mind the cost considerations. This total project cost is under $25 and if manufactured in bulk the cost will be even less.

## X. APPLICATION

A. Scientific Use

Recently, unmanned vehicles have found various scientific uses in hazardous and unknown environments. A lot of the probes which are sent to the other planets in our solar system have been unmanned vehicles, although some of the more recent ones were partially autonomous. The sophistication of these devices has fueled greater debate on the need for manned spaceflight and exploration. For example, we see that the Voyager I spacecraft is the first craft of any kind to leave the solar system. The Martian explorers Spirit and Opportunity have provided us with continuous data about the surface of Mars since January 3, 2004.The efficiency and effectivity of these vehicles are also quite appreciable.

B. Military and Law Enforcement Use

Military usage of remotely controlled military vehicles dates back to the first half of 20th century, when the Soviet Red Army used remotely controlled Tele tanks during 1930s in the Winter War and early stage of World War II. There were also remotely controlled cutters and experimental remotely controlled planes in the Red Army. Exposure to hazards is mitigated to the person who operates the vehicle from a location of relative safety. Remote controlled vehicles are used by many police department bomb-squads to defuse or detonate explosives. Unmanned Surface Vehicles (USVs) have undergone a dramatic evolution in capability in the recent past. Early USV's were capable of reconnaissance missions alone and but only with a limited range. Current USV's can hover around possible targets until they are positively identified before releasing their payload of weaponry. Use of USVs described in this paper can considerably improve upon the current structure.

C. Search and Rescue

USVs will also likely play an increased role in search and rescue missions. Slowly and steadily, all the developed nations (even some developing countries) are thinking about switching over and making use of these vehicles in case of natural disasters & emergencies. This can also be a great asset to save lives of both people along with soldiers in case of terrorist attacks like the one happened in 26 Nov, 2008 in Mumbai, India. The loss of military personnel can be largely reduced by using these advanced vehicles. This was demonstrated by the successful use of USVs during the 2008 hurricanes that struck Louisiana and Texas.

D. Forest Conservation

In the recent times, the wildlife has been prone to serious endangerment. Many animals are on the verge of becoming extinct, including the tiger to name a few. The spy robotic car can really be of help to us in this purpose. Since it is a live streaming device and also mobile, it can keep the forest guards constantly updated and alerted about the status of the different areas of the forest which are prone to attack. As a result, it can help to prevent further destruction of the forest resources and the wildlife by enabling correct prohibitory action at the appropriate time.

## XI. FUTURE SCOPE

The design procedure of an unmanned surface vehicle as presented in this paper can be further extended to include IR sensors and also a system to include password protection for the USV.IR sensors can be used to automatically detect & avoid obstacles if the vehicle goes beyond line of sight. This avoids damage to the vehicle if we are maneuvering it from a distant place. Project can be modified in order to password protect the vehicle so that it can be operated only if correct password is entered. Either cell phone should be password protected or necessary modification should be made in the assembly language code. This introduces conditioned access and increases security to a great extent. In case the vehicle is used as a spy car, a camera can also be mounted on the car. Such basic improvements can be made on the existing system as and when the requirement arises without making any major changes in the principle design of the USV.

## XII. CONCLUSION

By developing such an unmanned surface vehicle, we have overcome the drawbacks of the conventionally used RF circuits. This RCV includes advantages such as robust control, minimal interference and a large working range. The car requires five commands for motion control. The remaining controls are available to serve purposes dependant on the area of application of the RCV. We have tried to reduce the circuit complexity and improve upon the human machine interface. The cost analysis of the project described in section VII of this paper clearly indicates a huge improvement in the cost expenditure of the production of these unmanned vehicles. Moreover handling these USVs does not require much skill on the part of the user. Even an ordinary person can maneuver these USVs without having to know much about the internal circuitry. In this way the cost involved in training people to use such USVs can also be saved.